\documentstyle[preprint,revtex]{aps}
\begin{document}
\draft
\begin{title}
There is no $R^3 X S^1$ vacuum gravitational instanton
\end{title}
\author{Niall \'O Murchadha and Hugh Shanahan\cite{HS}}
\begin{instit}
Physics Department, University College, Cork, Ireland
\end{instit}
\begin{abstract}
Gravitational instantons, solutions to the euclidean Einstein equations, with
topology $R^3 XS^1$ arise naturally in any discussion of finite temperature
quantum gravity. This Letter shows that all such instantons (irrespective of
their interior behaviour) must have the same asymptotic structure as the
Schwarzschild instanton. Using this, it can be shown that if the Ricci tensor
of
the manifold is non-negative it must be flat. One special case is when the
Ricci tensor vanishes; hence one can conclude that there is no
nontrivial vacuum gravitational instanton. This result has uses both in
quantum and classical gravity. It places a significant restriction on the
instabilities of hot flat space. It also can be used to show that any static
vacuum Lorentzian Kaluza-Klein solution is flat.
\end{abstract}
\pacs{PACS numbers:04.60.+n, 04.50.+h}

\narrowtext

Instantons, solutions to the classical euclideanized field equations, play a
prominent role in quantum field theory. One situation in which they arise is in
computing a transition amplitude in the standard, Lorentzian, signature. The
transition amplitude may be dominated by a clasical solution to the field
equations. If such does not exist, it may be possible to deform the contour of
integration into a region of imaginary time and find that the integral may be
dominated by a solution to the euclidean field equations, an instanton. Such an
instanton may be interpreted as a tunneling solution.

Another use of instantons, of much more relevance to this Letter, is in finite
temperature quantum field theory. It can be shown that the partition function
at some given temperature is equivalent to a transition amplitude in which the
time is made both imaginary and periodic, with period $\tau = \beta = 1/T$,
where T is the temperature \cite{RF}. Again, the partition function may be
dominated by classical solutions, but now the classical solutions which one
considers are periodic as well as being euclidean,

Finite temperature quantum gravity has been intensively studied ever since the
discovery of black hole thermodynamics by Beckenstein\cite{JB} and
Hawking\cite{SH}. As part of this investigation, people have tried to find
gravitational instantons, especially those with a periodic character. It is
clear that one can identify flat slices of flat euclidean four-space to give a
flat instanton. Further, it was realised that if one took the standard
Schwarzschild solution and euclideanized it by $t \rightarrow i\tau$, one got a
regular vacuum instanton if one simultaneously made it periodic with period
$\tau_0 = 8\pi M$ \cite{GH}.

The Schwarzschild instanton has topology $R^2 X S^2$. It is widely assumed that
there is no vacuum gravitational instanton with topology $R^3 X S^1$ (except
rolled up flat space, of course). Such an instanton would signal an instability
of hot flat space \cite{GPY}. Here we will give a proof that no such instanton
exists. Edward Witten\cite{EW} has already shown that there is no nontrivial
vacuum gravitational instanton on $R^4$. The technique we use can be thought of
as an adaptation of the Witten proof.

 We assume that the metric is asymptotically flat in the $R^3$ directions. We
further assume that the the manifold has a constant period ($\tau_0$) near
infinity, but we do {\bf not} assume that the period remains constant in the
interior. In other words, we are assuming a constant temperature at infinity
but
we not care what happens in the interior.

The first point to be resolved is the asymptotic behaviour of Ricci-flat
Riemannian metrics on $R^3 X S^1$. Since the metric is asymptotically flat, we
can write the Einstein equation in the `Lorentz' gauge\cite{LL}. This means
that
we consider
\begin{equation}
h_{\mu\nu} = g_{\mu\nu} - \delta_{\mu\nu}
\end{equation}
and reverse its trace by defining
\begin{equation}
\tilde{h}_{\mu\nu} = h_{\mu\nu} - {1 \over 2} \delta_{\mu\nu}
g^{\alpha\beta}h_{\alpha\beta}\;.
\end{equation}
The `Lorentz gauge' means making a coordinate transformation so that
 $\tilde{h}$ satisfies
\begin{equation}
\tilde{h}_{\mu\nu,\nu} = 0\;.\label{2}
\end{equation}
Such a transformation can always be made. The linearized Einstein equations in
the Lorentz gauge is simply
\begin{equation}
^{(4)}\nabla^2 \tilde{h}_{\mu\nu} = 0\;.\label{3}
\end{equation}
Eq.\ (\ref{3}) holds only in the weak-field region, but this is the part of
the space where we assume that we have a constant period, $\tau_0$. We can
write  $\tilde{h}_{\mu\nu}$ as a Fourier series in $u$, the periodic
coordinate,
\begin{equation}
\tilde{h}_{\mu\nu} = \sum_{n=-\infty}^{\infty}\phi_{\mu\nu n}({\bf r})
e^{in\omega u} \label{4}
\end{equation}
where $\omega = 2\pi/\tau_0$ and ${\bf r}$ is the spatial radius vector. We
substitute Eq.\ (\ref{4}) into (\ref{3}) to get
\FL
\begin{eqnarray}
^{(4)}\nabla^2 \tilde{h}_{\mu\nu}& =& ^{(3)}\nabla^2 \tilde{h}_{\mu\nu} +
{\partial^2 \tilde{h}_{\mu\nu} \over \partial u^2}\\
& =&\sum_{n=-\infty}^{\infty} [^{(3)}\nabla^2\phi_{\mu\nu n}({\bf r}) - n^2
\omega^2\phi_{\mu\nu n}({\bf r})]e^{in\omega u}\\
& =& 0\;.\label{5}
\end{eqnarray}
Since the  Fourier components are linearly independent, Eq.\ (\ref{5}) implies
that each mode must satisfy
\begin{equation}
^{(3)}\nabla^2\phi_{\mu\nu n}({\bf r}) - n^2
\omega^2\phi_{\mu\nu n}({\bf r}) = 0\; ,\:\forall\; n\; .\label{6}
\end{equation}
It can be shown that the solutions of Eq.\ (\ref{6}) for $n \neq 0$ decay
exponentially at infinity. Therefore, the asymptotic behaviour is dominated by
the $n = 0$ mode which satisfies
 \begin{equation}
^{(3)}\nabla^2\phi_{\mu\nu 0} = 0\;.
\end{equation}
This means that the $n = 0$ mode is determined by the harmonic functions of the
flat-space three-laplacian. Hence the leading term is of the form
$C_0$/r, where $C_0$ is a constant. More precisely
\begin{equation}
\tilde{h}_{\mu\nu} \simeq {C_{\mu\nu} \over r}
\end{equation}
near infinity, where $C_{\mu\nu}$ are ten constants and $r$ is the three
dimensional radial distance. All the terms with $u$ dependence fall off
exponentially.

However, we must simultaneously satisfy the Lorentz condition (Eq.\ (\ref{2})).
This implies
\begin{equation}
{\partial \over \partial x^{\mu}}({C_{\mu\nu} \over r}) = 0\;.
\end{equation}
Therefore all the $C_{\mu\nu}$ 's must vanish except $C_{00}$, because
$r$ depends only on the three asymptotically flat coordinates (x, y. z).
Hence near infinity we have
\begin{eqnarray}
\tilde{h}_{\mu\nu} \simeq \;{C_{00} \over r}\left(\begin{array}{cccc}
1&\;0&\;0&\;0\\ 0&\;0&\;0&0\\ 0&\;0&\;0&\;0\\ 0&\;0&\;0&\;0 \end{array} \right)
\;.\label{13} \end{eqnarray}
We can reverse Eq.\ (\ref{2}) to give
\begin{equation}
h_{\mu\nu} = \tilde{h}_{\mu\nu} - {1 \over 2} \delta_{\mu\nu}
g^{\alpha\beta}\tilde{h}_{\alpha\beta}\;.\label{14}
\end{equation}
When we substitute (\ref{13}) in (\ref{14}) we get
\begin{eqnarray}
h_{\mu\nu} \simeq \;{C_{00} \over 2r}\left(\begin{array}{cccc}
1&\;0&\;0&\;0\\ 0&\;-1&\;0&0\\ 0&\;0&\;-1&\;0\\ 0&\;0&\;0&\;-1 \end{array}
\right) \;.\label{15} \end{eqnarray}
Finally, using $ -4M = C_{00}$, we get
\FL
\begin{eqnarray}
g_{\mu\nu} \simeq \left(\begin{array}{cccc}
1 - {2M \over r}&0&0&0\\ 0&1 + {2M \over r}&0&0\\ 0&0&1 + {2M \over r}&0\\
0&0&0&1 + {2M \over r} \end{array} \right)  + {\cal O}({1 \over
r^2})\;.\label{16}
\end{eqnarray}
We should recognise Eq. (\ref{16}) as being just the leading part of the
Schwarzschild instanton\cite{GH}. This should come as no surprise because all
we
have been doing is determining the asymptotic behaviour of the
gravitational instanton, and $R^3 X S^1$ is indistinguishable from $R^2 X S^2$
near infinity. This result holds even if the Ricci tensor of the manifold is
non-zero; all we require is that it fall off sufficiently rapidly at infinity.

 This means that if we wish to obtain a nonexistence result we need a global
argument, a `local near infinity' argument will never get us anything. This is
why we try to mimic the Witten $R^4$ proof\cite{EW}.

Following Witten, we seek a solution to the equation
\begin{equation}
^{(4)}\nabla^2\phi^0 = 0\;,\; \phi^0 \rightarrow u \hbox{ near infinity}.
\label{17}\end{equation}
Let us write
\begin{equation}
\phi^0 = u  + \phi^{(0)}_1\;.
\end{equation}
Eq.\ (\ref{17}) can be written as
\begin{equation}
^{(4)}\nabla^2\phi^{(0)}_1 = -^{(4)}\nabla^2 u\;,\; \phi^{(0)}_1 \rightarrow
0 \hbox{ near infinity}. \label{19}
\end{equation}
We have
\begin{equation}
^{(4)}\nabla^2 u = -g^{\alpha\beta}\Gamma^u_{\alpha\beta} = {\cal o}({1 \over
r^3})\;,
\end{equation}
which means that a solution to (\ref{19}) exists which decays at infinity at
least as fast as 1/$r$, and hence we have a solution to (\ref{17}).

The following identity is now used:
\FL
\begin{equation}
\nabla_{\mu}(\nabla_{\nu}\phi^0\nabla^{\nu}\nabla^{\mu}\phi^0) =
(\nabla_{\nu}\nabla_{\mu}\phi^0)^2 + \nabla_{\nu}\phi^0
\nabla_{\mu}\nabla^{\nu}\nabla^{\mu}\phi^0\;.\label{21}
\end{equation}
Let us add and subtract $\nabla_{\nu}\phi^0
\nabla^{\nu}\nabla_{\mu}\nabla^{\mu}\phi^0$ (which is identically zero from
(\ref{17})) to Eq.\ (\ref{21}) to give
\FL
\begin{equation}
\nabla_{\mu}(\nabla_{\nu}\phi^0\nabla^{\nu}\nabla^{\mu}\phi^0) =
(\nabla_{\nu}\nabla_{\mu}\phi^0)^2 + \nabla_{\nu}\phi^0\nabla_{\mu}\phi^0 R
^{\mu\nu}\label{22}
\end{equation}
Since we assume that the manifold is Ricci-flat, we can throw away the last
term in Eq.\ (\ref{22}). Let us now integrate (\ref{22}) over the whole
manifold to give
\FL
\begin{equation}
\int\nabla_{\mu}(\nabla_{\nu}\phi^0\nabla^{\nu}\nabla^{\mu}\phi^0)\sqrt{g} d^4x
= \int(\nabla_{\nu}\nabla_{\mu}\phi^0)^2\sqrt{g} d^4x \label{23}
\end{equation}
The left-hand-side of (\ref{23}) can be turned into a surface integral at
infinity to give
\FL
\begin{equation}
\oint_{\infty}\nabla_{\nu}\phi^0\nabla^{\nu}\nabla^{\mu}\phi^0\hat{n}_{\mu}\sqrt
   {g}
d^3S = \int(\nabla_{\nu}\nabla_{\mu}\phi^0)^2\sqrt{g} d^4x \label{24}
\end{equation}
Any term in the integrand of the surface integral that falls off faster than
$1/r^2$ can be neglected because the `area' of the `surface at infinity' blows
up like $4\tau_0 \pi r^2$ where $\tau_0$ is the period in the $u$ direction.
Thi
   s
allows us to ignore the contribution from the $\phi^{(0)}_1$ term. The only
term
that remains is the connection from $\nabla^{\nu}\nabla^{\mu}\phi^0$. It is
easy to show that the surface integral reduces to
\begin{equation}
-4\pi \tau_0 \Gamma^u_{ur} = -4\pi M\tau_0\;.
\end{equation}
Hence (\ref{24}) gives us
\begin{equation}
\int(\nabla_{\nu}\nabla_{\mu}\phi^0)^2\sqrt{g} d^4x = -4\pi
M\tau_0\;.\label{26}
\end{equation}
Therefore $M$, the analogue of the Schwarzschild mass, must be negative.

However, $\phi^0$ is not the only natural harmonic function we could define on
this manifold. Another candidate is
\begin{equation}
^{(4)}\nabla^2\phi^1 = 0\;,\ \phi^1 \rightarrow x \hbox{ near infinity}.
\label{27}\end{equation}
We repeat the calculation following (\ref{17}) and write
\begin{equation}
\phi^1 = x  + \phi^{(1)}_1\;,
\end{equation}
and we again have
\begin{equation}
^{(4)}\nabla^2 x = -g^{\alpha\beta}\Gamma^x_{\alpha\beta} = {\cal O}({1 \over
r^3})\;.\label{29}
\end{equation}
The connection, in general, falls off like 1/$r^2$, but a cancellation occurs
in the particular combination in (\ref{29}) so that the leading terms cancel.
In other words, the standard coordinates on any manifold which is
asymptotically Schwarzschildian are `almost harmonic'.

We can use exactly the same identity as before, just substituting $\phi^1$ for
$\phi^0$. Now we get, instead of (\ref{26}),
\begin{equation}
\int(\nabla_{\nu}\nabla_{\mu}\phi^1)^2\sqrt{g} d^4x = -4\pi r^2
\Gamma^x_{xr} = +4\pi M\tau_0\;.\label{30}
\end{equation}
Now we get $M > 0$. But we have already shown $M < 0$!

 The only way that (\ref{30}) can be compatible with (\ref{26}) is that we
really have $M = 0$, and this implies
\begin{equation}
 \nabla_{\nu}\nabla_{\mu}\phi^0 = \nabla_{\nu}\nabla_{\mu}\phi^1 = 0 \;.
\end{equation}
The existence of these functions (and their equivalents $\phi^2$ and $\phi^3$)
whose double derivatives vanish is sufficient to show that the 4-space is flat.
This non-existence argument extends to the case where we have a non-zero Ricci
tensor if the Ricci tensor has non-negative eigenvalues (see \cite{JW} for a
similar result in the $R^4$ case).

The undoubted existence of the Schwarzschild instanton does not contradict this
non-existence result. It is very easy to repeat this calculation in the
Schwarzschild case because the standard time coordinate $t$ in the
Schwarzschild
instanton is harmonic due to the static nature of the metric. When we evaluate
expression (\ref{24}), it turns out that we should regard the manifold as
having an `interior' surface as well as the surface at infinity. Even though
the area of the `surface' at $r = 2M$ is zero,  the integral of $\nabla_{\nu}t
\nabla^{\nu}\nabla^{\mu}t \hat{n}_{\mu}$ actually diverges on any surface
that shrinks to that point. Thus the negative term at infinity is more than
compensated for by a positive interior term. We assume that our manifold is
globally regular so no such interior surfaces need be allowed for.

Over the years, a number of `static + vacuum implies trivial' theorems have
been
derived (see, for example, \cite{SD}). (The only counterexample to date has
been
the Einstein-Yang-Mills system \cite{BM}.) The result obtained here, that a
Ricci-flat Riemannian manifold with topology $R^3 X S^1$ is flat, can be used
to
show that `static + vacuum implies trivial' is valid for standard (Lorentzian)
Kaluza-Klein theory. In Kaluza-Klein theory we consider a manifold with
topology
$R^4 X S^1$ where the $S^1$  and three of the four directions in $R^4$ are
spacelike. Let us consider a static, vacuum Kaluza-Klein manifold. By static we
mean that there exists of a timelike, surface-forming Killing vector.
Obviously,
the surface orthogonal to the Killing vector, $M^4$, is a Riemannian manifold
with topology $R^3 X S^1$. The `Einstein' equations in this case reduce to
\begin{equation}
N R_{\mu\nu} = \nabla_{\mu}\nabla_{\nu} N,\;\; ^{(4)}\nabla^2 N = 0
\;,\label{32}
 \end{equation}
where $R_{\mu\nu}$ is the Ricci tensor of the four-manifold, and $N$ is the
norm of the Killing vector. The second equation in (\ref{32}) tells us that $N$
must be constant, the first equation then tells us that the Ricci tensor must
vanish. Therefore $M^4$ must be flat and hence the five-manifold must be
trivial.
\acknowledgments
We would like to thank Michel Vandyck, especially for his help with the proof
that the zero mode dominates. We would also like to thank Raphael Sorkin, who
indicated the Kaluza-Klein result to one of us (N\'OM) .

 \end{document}